# Doers, not Watchers: Intelligent Autonomous Agents are a Path to Cyber Resilience

Alexander Kott, US Army CCDC Army Research Laboratory, USA

Paul Theron, Thales Aerospace Cyber Resilience research chair, France

Today's cyber defense tools are mostly watchers. They are not active doers. To be sure, watching too is a demanding affair. These tools monitor the traffic and events; they detect malicious signatures, patterns and anomalies; they might classify and characterize what they observe; they issue alerts, and they might even learn while doing all this. But they don't act. They do little to plan and execute responses to attacks, and they don't plan and execute recovery activities. Response and recovery -- core elements of cyber resilience (Kott and Linkov 2019) -- are left to the human cyber analysts, incident responders and system administrators. A recent report (NAS 2019) reviews the implications of AI for cyber security and offers no examples of applications of AI to response and recovery.

We believe things should change. Cyber defense tools should not be merely watchers. They need to become doers -- active fighters in maintaining a system's resilience against cyber threats. This means that their capabilities should include a significant degree of autonomy and intelligence for the purposes of rapid response to a compromise – either incipient or already successful – and rapid recovery that aids the resilience of the overall system. Often, the response and recovery efforts need to be undertaken in absence of any human involvement, and with an intelligent consideration of risks and ramifications of such efforts. Recently an international team (the NATO Science and Technology Organization's Research Group IST-152) published a report that proposes a vision of an autonomous intelligent cyber defense agent (AICA) and offers a high-level reference architecture of such an agent (Kott et al. 2019). Let's explore this vision.

## It's the agents' world

What motivates the envisioned, admittedly ambitious, capabilities of AICA? Why are they important? To explain, let us begin with a domain where needs for such active agents are most salient and perhaps less controversial – the domain of warfare. The battlefield of the future will be crawling with intelligent agents of various sorts. Some will be physical robots while others will be disembodied, software agents. The proliferation of intelligent agents is the

APPROVED FOR PUBLIC RELEASE

already emerging reality of warfare. In the near future, they will form an ever-growing fraction of total military assets (Scharre 2014) – many more agents (robots or software) than humans.

Some of these agents will be sophisticated cyber attackers, and all of them will be potential, lucrative targets of cyber-attacks. Highly capable and pervasive malware – a type of cyber agents – will continually penetrate and operate on the opponent's devices. Every combatant of the future battlefield must assume that his systems have a high probability of being compromised and infested with the adversarial cyber agents. In a world dominated by multiple ubiquitous intelligent agents, cyber warfare is pervasive and hyperactive: the agents are both the targets and the perpetrators of the attacks; they make this form of warfare both profitable and feasible.

In such a world, the reliance on human cyber defenders and responders will be untenable. Indeed, locally on the battlefield, the number of agents – both friendly and adversarial – will far exceed the number of human warfighters who are available and capable of performing cyber defense tasks. The cognitive demands imposed on the humans in such a complex environments would be unmanageable. Besides, the overall tempo of operations will be such that human defenders' reaction time will be inadequate in most cases (Kott 2018).

Another possibility might be to rely on centralized monitoring and defense services. Much of today's cyber security relies on collecting cyber-relevant data from multiple devices, submitting them to a central, well-equipped and staffed location, which then performs the appropriate analysis and response actions. This approach, however, is of limited utility on the future battlefield, for several reasons. First, all communications on the battlefield come with great risk: the adversary uses such communications to locate and destroy the communicating entities. Second, communications are jammed and may not be possible for any prolonged or predictable periods. Third, once a capable malware penetrates a device, it seeks to disable or spoof the means by which a device can obtain help from an external monitoring center.

So, let's recap the options. If – or rather when -- a malware compromises a computer of a manned tank, for example, the soldiers operating the tank will unlikely have either time or skills to deal with the compromise. The same applies if the tank is unmanned, and even if a human soldier happens to be nearby. Finally, a remote support option is infeasible in an environment where communications are disabled by cyber or electromagnetic means. The only remaining option is to have an artificial cyber defender residing locally on the computer. And this artificial defender must be intelligent enough to plan and execute decisive actions, autonomously, in order to keep the computer performing its mission in spite of the compromise.

Do these arguments apply only to extreme conditions of warfare? No. The domain of active military operations is hardly the only domain that calls for faster and more assertive behavior of autonomous cyber defenders. Even today, in civilian environments, cyberattacks are fast, brutal, highly consequential, and demand extremely fast response. Criminals or irresponsible pranksters are able to take control of cars traveling at high speed, or planes in the air that may constitute mortal threat to their passengers and other innocent victims (Ring 2015). An onboard intelligent autonomous agent capable of taking the necessary response and recovery actions is needed in such civilian environments no less than on a battlefield.

## How smart does it need to be?

Returning to the example of AICA agent that resides on a computer of an unmanned tank – where we suspect an adversary's malware operates already -- let's consider the capabilities the agent must possess. Fig 1. depicts the functional components of the agent, partly derived from the widely accepted model of Russell and Norvig (2009).

Using its connections to sensors and actuators (both physical and cyber), the agent collects information about the processes and events occurring in the computer, and uses this information to infer and characterize the state of the system, including the malware if any. It should be emphasized that all such information collection – and any resulting actions -- has to be performed in a stealthy mode. The agent must minimize its visibility to the malware that is likely to lurk within the same computer. In general, AICA must plan and execute all its actions in a way that minimizes the probability that the adversary malware will detect and attack the agent.

As the state of the system changes, AICA keeps track of the changes and analyzes their significance. If the trajectory of changes indicates potentially undesirable ramifications, the agent formulates a plan (or multiple alternative) plans of actions, estimates or simulates the effects of executing the actions, and selects the next most appropriate actions to execute, in the context of the preferred plans. The actions may modify or strengthen the system, or may serve to degrade the effectiveness of the malware.

In most cases, AICA performs this decision making without having the luxury of consulting a human operators or analysts, either because the time is too short or because the human cannot be contacted. Nevertheless, provisions are available to consult a human operator when feasible. The agent may also collaborate with agents residing on other vehicles, or other computers within the same vehicle. However, because all communications are likely impaired or observed by the adversary, in most cases the agent operates alone.

With such a self-reliance, a degree of real-time learning is necessary, particularly because the tactics, techniques and procedures of the adversary's malware evolve rapidly. AICA may need to learn at least to differentiate between normal and anomalous patterns in its environment, and perhaps learn to recognize when its actions fail to achieve results under certain conditions. More extensive types of machine learning may be reserved to a centralized controller of multiple AICA agents, which will update the agents when communications become available and safe.

To fight the adversary malware, or the effects of that malware's depredations, AICA often has to take destructive actions or modifications, such as deleting, quarantining, or reconfiguring certain software. Such potentially risky actions are performed only on the computer where AICA resides, and even then under strict rules. The agent uses its plan generation and simulation functions to anticipate and minimize such risks.

Still, there can be no guarantee of preserving the availability or integrity of the functions and data of the computer. Inevitably, a risk exists that an action of AICA will "break" the computer, disabling a critical software, or corrupting important data. This risk, in a military environment, has to be balanced against the death or destruction caused by the adversary if

APPROVED FOR PUBLIC RELEASE

the agent's action is not taken. In many cases, similar balance applies to civilian applications as well.

Should AICA be a mobile agent? It is possible to envision that an agent will move (or move a replica of itself) to another friendly computer. We envision only very restricted conditions –and perhaps none at all -- under which such a propagation may occur. Doing otherwise would needlessly open a yet another round of unproductive controversy about "good viruses" (Muttik 2016).

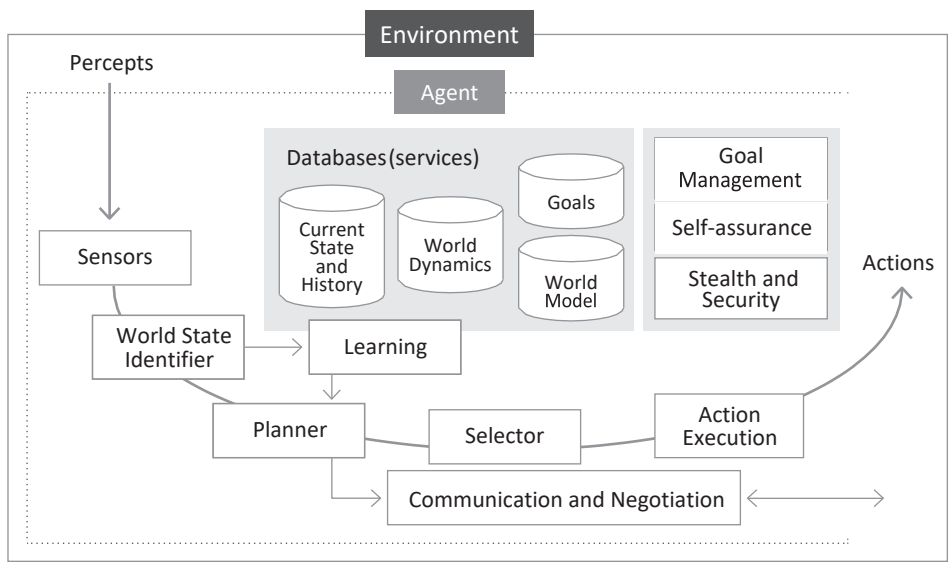

Figure 1. Functional Architecture of an Autonomous Intelligent Cyber Agent (Kott et al 2018).

## Engineering and research challenges

AICA is an ambitious vision, and we are yet to see an example that approximates a majority of the envisioned capabilities. Engineering challenges inherent in such an agents include the complexity of design; quality assurance and certification; testability and simulation of the artifact; and achieving acceptable performance within the constraints of available computing power. Other challenges point to a potentially extensive agenda of research and development of the fundamental functions of AICA.

The agent will require a rich and comprehensive representation (Müller 2007) and modeling of the complex world surrounding the agent, of its own capabilities and resources, of the processes and techniques inherent in responding to the hostile actions and counter-action, and in recovering from such actions, and of the knowledge required to reason and make decisions with respects to all these information. Approaches are needed to achieve such representations that are sufficiently powerful and yet compact and affordable in terms of building and maintaining them over the life cycle of the agent.

Affordable formation and adaptation of AICA's knowledge bases will likely rely on machine learning (Ridley 2018), especially as the adversary's behaviors and techniques, as well as the environment's features and vulnerabilities will evolve rapidly. The appropriate methods of machine learning, and how it should be divided between onboard and off-board learning, and how much of the learning should occur in real-time are major topics for research.

Some of the challenges involve limitations of onboard computing resources (for onboard, real-time learning) and limitations of communications bandwidth (for off-board learning). With highly fluid environments and adversaries, learning (and validation of the results) with a small number of samples – a challenging topic of ongoing research – will be necessary. Some of the samples will be a product of intentional deception by the adversary. In general, issues of Adversarial Learning (Papernot et al., 2016) are of great importance to ML.

AICA's decision-making processes, and its quality, are the key to the agent's effectiveness and trustworthiness, and remain the subject of relatively early research (Heinl 2014). Given its essentially adversarial nature, it must involve elements of adversarial reasoning and game theoretic strategy analysis. Integration of several decision-making approaches is likely to be required, in particular incorporating a degree of decision-making plasticity as an adaptive response to characteristics and uncertainty of a particular situation (Theron 2014).

Although we stressed the importance of AICA's ability to act alone, autonomously, it does not mean that the agent should not take advantage of opportunities to collaborate, when possible, with a central C2 node, a human operator, or with another, friendly intelligent agent. Such a collaboration may improve quality and certainty of the agent's action plan, engage collaborative actions of other agents, and in general strengthen the ability of AICA to provide effective response and recovery. Research is needed on collaborative reasoning, protocols, resilience, and trustworthiness of such collaborations.

In particular, considerations of AICA's stealth put significant constraints on collaboration. Any communications with another entity, either within a given computer or with another device, have a chance of being noticed by an adversary. AICA will be a high-priority target for a hostile malware, which will seek to find and disable the agent. Approaches to deciding whether another entity is a friend or foe – a key element of a decision to engage in collaboration -- is yet another research challenge associated with the overall goal of minimizing observability and maximizing survivability of AICA.

## Current prototypes

Is it even possible to build such an agent? Are we asking for too much? Perhaps not. To be sure, we have not seen a prototype of an agent that embodies even a plurality, let alone all, functions envisioned in AICA. However, prototypes that illustrate certain functions of such an agent do begin to appear.

For example, the AHEAD architecture (de Gaspari et al. 2016) defines an autonomous agent equipped with a set of active defense tools that dynamically change their configuration to implement the decisions taken by the controller. The AHEAD system provides many functions envisioned in the AICA architecture, such as (Fig. 1) Sensing and World State Identification, Planning and Action Selection, Action Execution, and Learning.

Deception is a highly effective action that an intelligent agent can take to defend a system against malware. Deception also offers the agent an opportunity to remain stealthy. In an example of a related research, an agent performs dynamic analysis of the detected malware and then plans and executes several types of deceptive actions depending on the behavior and intents of the malware. The malware remains unaware that it is being deceived (Al Shaer et al. 2019).

## A path forward

Our NATO research group envisioned three phases in a roadmap towards a full implementation of an AICA. The first phase would develop a knowledge-based planning of actions, the execution functionality, and initial capabilities of the agent to operate resiliently while under attack. Then, in a series of Turing-like experiments, developers would evaluate the capability of the agent to produce plans of remediating a compromise, by comparing the plans to those of experienced human cyber defenders.

The second phase would focus on adaptive learning, the development of a structured world-model, and mechanisms for dealing with explicitly defined, multiple and potentially conflicting goals. Here, the prototype agent should demonstrate the capability, in a few self-learning attempts, to return the defended system to acceptable performance after a significant change in the adversary behavior or techniques. Finally, the third phase would delve into issues of multi-agent collaboration, human interactions, and ensuring both the stealth and trustworthiness of the agent.

It would be highly desirable to see formation of an interest group that could give impetus toward development of AICA-like generation of agents. Such a group would certainly center on industry members but also include academia and perhaps governments. Together, they could encourage development of a variety of commercial agent products, by providing common intellectual underpinnings, e.g., recommendations for terminology, requirements, standards, and test cases.

Acknowledgement: The authors are grateful to the members of the NATO Science and Technology Organization's Research Group IST-152.

Disclaimer: This article does not reflect the positions or views of the authors' employers.

## References

Al-Shaer E, Wei J, Hamlen KW, Wang C. Towards intelligent cyber deception systems. In: Autonomous cyber deception: reasoning, adaptive planning, and evaluation of honeythings. New York (NY): Springer; 2019.

C. H. Heinl, "Artificial (Intelligent) Agents and Active Cyber Defence: Policy Implications," in 6th International Conference on Cyber Conflict, P. Brangetto, M. Maybaum and J. Stinissen,

This is a pre-print version of: Kott, A. and Theron, P., 2020. Doers, Not Watchers: Intelligent Autonomous Agents Are a Path to Cyber Resilience. *IEEE Security & Privacy*, *18*(3), pp.62-66.

Eds., Tallinn, NATO CCD COE Publications, 2014, pp. 53-66.

De Gaspari F, Jajodia S, Mancini LV, Panico A. AHEAD: a new architecture for active defense. In: Proceedings of the 2016 ACM Workshop on Automated Decision Making for Active Cyber Defense (SafeConfig); 2016.

Kott, Alexander. Ground Warfare in 2050: How It Might Look. No. ARL-TN-0901. US Army Research Laboratory Aberdeen Proving Ground United States, 2018.

Kott, A. and Linkov, I. eds., 2019. *Cyber resilience of systems and networks*. Springer International Publishing.

Kott, A., P. Theron, M. Drašar, E. Dushku, B. LeBlanc, P. Losiewicz, A. Guarino, L. V. Mancini, A. Panico, M. Pihelgas and K. Rzadca, "Autonomous Intelligent Cyber-defense Agent (AICA) Reference Architecture, Release 2.0," US Army Research Laboratory, Adelphi, MD, 2019.

Muttik I. Good viruses. Evaluating the risks. DEF CON 16; 2008 Aug 8–10; Las Vegas, NV. https://www.defcon.org/images/defcon-16/dc16-presentations/defcon-16-muttik.pdf

Müller, V.C., 2007. Is there a future for AI without representation?. *Minds and Machines*, *17*(1), pp.101-115.

[NAS] National Academies of Sciences, Engineering, and Medicine 2019. Implications of Artificial Intelligence for Cybersecurity: Proceedings of a Workshop. Washington, DC: The National Academies Press. https://doi.org/10.17226/25488.

Papernot, N., P. McDaniel, S. Jha, M. Fredrikson, Z. Celik, and A. Swami, The limitations of deep learning in adversarial settings. In Security and Privacy (EuroS&P), 2016 IEEE European Symposium, 372-387.

A. Ridley, "Machine learning for Autonomous Cyber Defense," The Next Wave, vol. 22, no. 1, pp. 7-14, 2018.

Ring, T., 2015. Connected cars – the next target for hackers. Network Security, 2015(11), pp.11-16.

S. Russell and P. Norvig, Artificial intelligence: a modern approach. 3rd ed., London (UK): Pearson, 2010.

Scharre, P. Robotics on the Battlefield Part II: The Coming Swarm, Report, Center for a New American Security, Washington, DC, 2014.

P. Theron, "Lieutenant A and the rottweilers. A Pheno-Cognitive Analysis of a fire-fighter's experience of a critical incident and peritraumatic resilience," PhD Thesis, University of Glasgow, Scotland, 2014. https://sites.google.com/site/cognitionresiliencetrauma

## About the Authors:

ALEXANDER KOTT is the Chief Scientist, US Combat Capabilities Development Command's Army Research Laboratory. Contact him at alexander.kott1.civ@mail.mil.

PAUL THERON is Thales Aerospace's Cyber Resilience (Cyb'Air) Research Chair, France. Contact him at paul.theron@thalesgroup.com.